# NEMAR: An open access data, tools, and compute resource operating on NeuroElectroMagnetic data

Arnaud Delorme, Dung Truong, Choonhan Youn, Subha Sivagnanam, Kenneth Yoshimoto, Russell A. Poldrack, Amit Majumdar, Scott Makeig

*Abstract—* To take advantage of recent and ongoing advances in large-scale computational methods, and to preserve the scientific data created by publicly funded research projects, data archives must be created as well as standards for specifying, identifying, and annotating deposited data. The OpenNeuro.org archive, begun as a repository for magnetic resonance imaging (MRI) data, is such an archive. We present a gateway to OpenNeuro for human electrophysiology data (BIDS-formatted EEG and MEG, as well as intracranial data). The NEMAR gateway allows users to visualize electrophysiological data, including time-domain and frequency-domain dynamics time locked to sets of experimental events recorded using BIDS- and HED-formatted data annotation. In addition, NEMAR allows users to process archived EEG data on the XSEDE high-performance resources at SDSC in conjunction with the Neuroscience Gateway (nsgportal.org), a freely available and easy to use portal to leverage high-performance computing resources for neuroscience research..

## I. INTRODUCTION

Sharing of electrophysiological data, as well as evaluating it by direct comparison with ever larger stores of data made available from previous projects, is critical to leveraging public research investment and to supporting rigor and reproducibility in funded research. Although electroencephalography (EEG) was the first functional human brain monitoring modality (1926-), EEG data analysis long lagged in adapting new data analysis approaches – in neurology, where visual pattern recognition is still the dominant approach, and in cognitive neuroscience, in which event-related potential (ERP) averaging of individual scalp channel records was long the dominant research method. These methods, however, cannot exploit consistencies in complex data that can only be identified in and extracted from large to very large data collections.

Here we report initial results in building a large human neuroelectromagnetic (here, NEM) data archive and tools resource (NEMAR). Our goal is to build a highly used open resource for the archiving, sharing, analysis, and mining of NeuroElectroMagnetic (NEM) imaging data.

Because high-density EEG, MEG, and iEEG recording have no in-common scalp sensor (electrode or SQUID coil) placements, as well as the important physical differences that exist between the dozens of EEG/iEEG and several available MEG systems, a centralized archive of NEM data allowing direct comparison of NEM brain dynamics across studies and modalities has not previously been feasible. Further, the broad point-spread of cortical magnetic flux to scalp coils, and the still broader and more variable point-spread of cortical potentials to scalp electrodes mean that scalp and intracranial NEM recordings have no simple relationship. As a result, much valuable convergent information about human brain dynamics contained in the many large and small existing and in-progress NEM data sets, each recorded with care at considerable expense, is at strong risk of being lost to science unless and until the data are integrated into an active integrated data, tools, and compute resource making possible advanced analysis within or across studies.

## II. DATA FORMATS

### A. The BIDS format

In the past few years, the MRI community have worked jointly with the International Neuroinformatics Coordinating Facility (INCF), to develop a community standard for describing and annotating MRI data, the Brain Imaging Data Structure (BIDS) [1]. To speed adoption by the brain imaging community, community-driven BIDS standards use common file formats (for MRI, NIFTI, JSON, TSV) and simple directory structures, and do not require additional database software. The BIDS MRI standard has now been adopted by several data repositories including OpenNeuro, FCP INDI, SchizConnect, and the Developing Human Connectome Project. The BIDS format for MRI has been responsible in large part for a recent rapid rise in data sharing in the MRI and fMRI world [2,3]. BIDS has gained rapid acceptance as an evolving, community-based standard for organizing neuroimaging data to enable efficient data search, advanced processing, and data mining. Benefits of the OpenNeuro approach to fMRI data sharing and computation are becoming more apparent, prompting adoption from the NEM data research community. BIDS extensions to MEG data [4] and, more recently, to electrophysiology data including EEG [5] and iEEG data [6] have been developed. Although several small and dedicated NEM data repositories exist (e.g., HeadIT.org, iEEG.org, the OMEGA MEG data archive), these archives do not build around a common data format standard, thwarting cross-modality data co-registration and federation. NEMAR's support of the emerging BIDS NEM-modality standards will allow storage of a wide variety of NEM data collection formats and protocols, and make identification, storage, and efficient search and retrieval of a wide variety

[1]*Research supported by NIH (5R24MH120037-02).

A. Delorme, D. Truong, and S. Makeig are with the Swartz Center of Computational Neuroscience, INC, UCSD, La Jolla, CA, USA (e-mail: arnodelorme@gmail.com, smakeig@ucsd.edu).

A. Delorme is also with CERCO, CNRS, Paul Sabatier University, Toulouse, France

A. Majumdar, K. Yoshimoto, C. Youn, S. Sivagnanam are with the San Diego Supercomputer Center, UCSD, La Jolla, CA, USA

R. Poldrack is with Stanford University, Stanford, CA, USA

of NEM data possible within a single archive or federated network of archives.

*B. The HED standard*

Archived time series data typically require that standardized terms be used to describe the natures of all experimental events of interest identified in the recording session, either during its capture or after the fact, for potential search and reuse in further analysis and meta-analysis. Although BIDS-formatted data sets contain detailed metadata, the BIDS standards in themselves do not constitute a system for adequately describing experimental events. The Hierarchical Event Descriptor (HED) system provides a standardized, flexible, and readily extensible set of descriptors for experimental events in brain imaging or behavioral experiments [7]. HED tagging can be used to describe many types of experiment events in a uniform but easily extensible, and both human- and machine-readable manner. HED has been integrated into BIDS as the *de facto* standard for describing events in brain imaging experiments. The NEMAR archive is the first open data archive to leverage the use of HED tags to enable data discovery and integration based on experimental events.

### III. IMPLEMENTATION

*A. NEMAR as an extension of the OpenNeuro project*

The OpenNeuro archive [8] currently offers more than 644 open neuroimaging datasets from more than 22,000 participants. Its success to date reveals that neuroscientists are willing to share their data, although the attractiveness of sharing remains limited as data sharing can require extra work -- not currently required by journals or funding agencies -- and across-studies data meta-analysis and mining tools are not yet available. The OpenNeuro architecture has three main components: front end, database. The custom web front end runs on an Amazon Web Services server. The data are stored in an AWS S3 instance, backed by the open-source Datalad data management software, which has a data snapshotting mechanism allowing OpenNeuro users to version their data.

The NEMAR project will capitalize on the established and ongoing achievements of the OpenNeuro, Neuroscience Gateway (NSG) [9], Open EEGLAB Portal [10], and BIDS standards development projects by creating a community portal to a large and ever-growing archive of human neuroelectromagnetic (NEM: EEG, MEG, iEEG) brain imaging data, data analysis tools, and advanced computational resources. Our overall goal is to support the creation, maintenance, analysis, and cross-study mining of human neuroelectromagnetic (NEM) data while seeding and growing 'minable' archives of NEM data in the OpenNeuro resource. After users upload NEM data to OpenNeuro, the data are automatically copied to SDSC storage using the DataLad cloning mechanism. New DataLad snapshots from OpenNeuro are synched daily. NEM relevant data measures are then automatically computed on the Neuroscience Gateway and displayed to users using the NEMAR interface.

The NEMAR website (*nemar.org*) uses the HUBzero web framework. HUBzero, maintained and developed at the San Diego Supercomputer Center (SDSC), is an open-source software platform for building powerful websites that host analytical tools, publish data, share resources, collaborate, and build communities in a single web-based ecosystem [13].

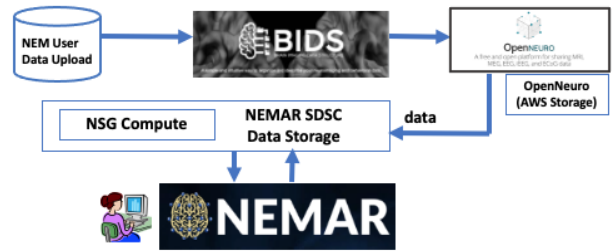

*Figure 1. Users upload their data to OpenNeuro. The data are then synced with NEMAR providing additional functionally (data visualization and on-site computation).*

*B. NEMAR BIDS-formatting tool*

We have released a *bids-matlab-tools* plug-in (version 6) for the popular EEGLAB [11] software environment (*sccn.ucsd.edu/eeglab*) running on MATLAB (The Mathworks, Inc.) to export EEG data to BIDS. The plug-in allows users to format their EEG raw data using the BIDS-EEG structure either via the BIDS export menu item, a pop-up window (Fig. 1), or using a scripting interface. The *bids-matlab-tools* plug-in, available from the EEGLAB plug-in manager, has been downloaded 844 times as of January 2022.

The plug-in guides users through a series of questions to describe their experiment and populate the BIDS metadata fields, as well as specify HED tags. Upon completion, the plug-in exports a fully compliant BIDS-formatted dataset that may be uploaded to the OpenNeuro using its web or command line interface, after which the data are automatically made available on NEMAR.

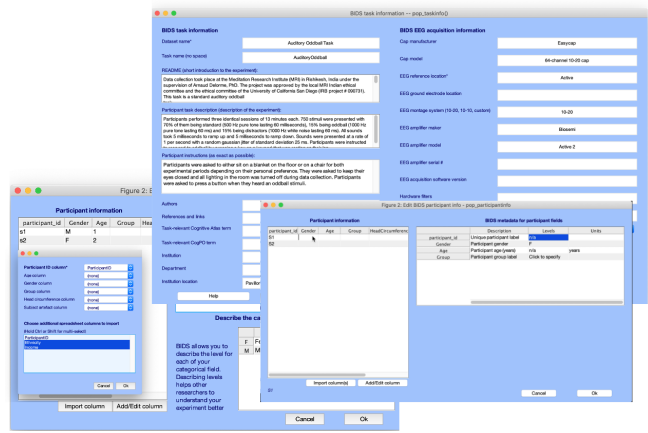

*Figure 2. Graphic interface of the bids-matlab-tools plug-in for EEGLAB, a tool to format NEM data to BIDS for NEMAR and OpenNeuro.*

*C. NEMAR dataset web interface and meta-data*

NEMAR automatically displays a variety of data information extracted from BIDS including the number of EEG channels (and other channel types), the size of the data, the number of files, the BIDS and HED versions, authors, dataset DOI (from OpenNeuro), and other BIDS related meta-data (Fig. 3).

As in OpenNeuro, users can explore the BIDS folder content and visualize and/or download data files using the icons adjacent to the data files.

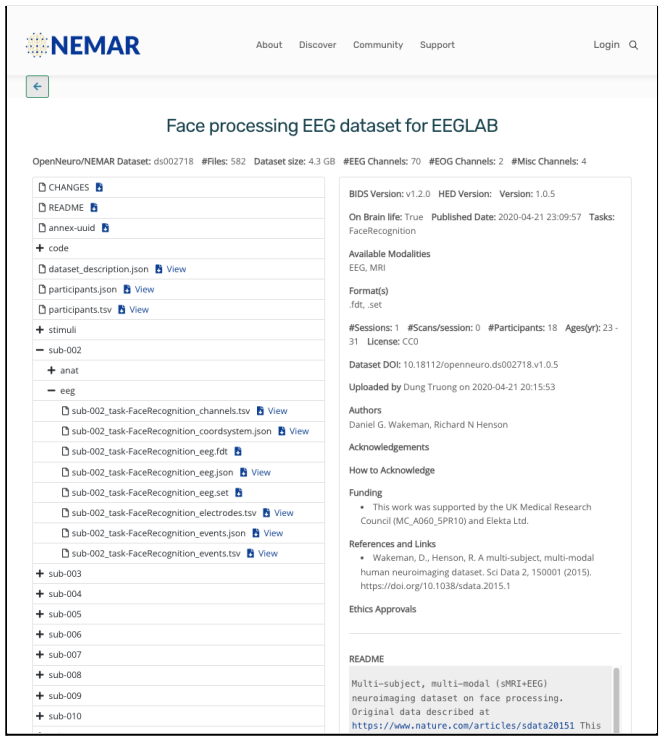

*Figure 3. NEMAR dataset interface exposing some of the BIDS required EEG metadata.*

Currently (02/22) there are 72 EEG datasets (from 2,664 participants) on NEMAR, 22 MEG datasets (from 365 participants) and 12 iEEG datasets (from 202 participants). Of the 72 EEG datasets, 37 use the EEGLAB data format and have likely been formatted using the *bids-matlab-tools* EEGLAB plug-in (see III.B).

*D. NEMAR data quality pipeline*

On NEMAR we are implementing an automated processing pipeline to compute a variety of data quality metrics, including the percentage of 'good' channels and 'good' data, and the number of putative brain sources extracted by independent component analysis, as described in [12]. These measures are shown when users click a link next to each dataset (Fig. 4).

For each data file of each BIDS dataset, NEMAR also plots the data spectrum and shows a few seconds of the raw data. More data-specific measures will be added in the future. Measures shown by NEMAR are computed using EEGLAB on NSG (see III.E). Visualization measures are saved in the .SVG vector format, and are displayed using native SVG HTML tag capabilities, preserving vector graphic information for adaptive resolution on web browsers.

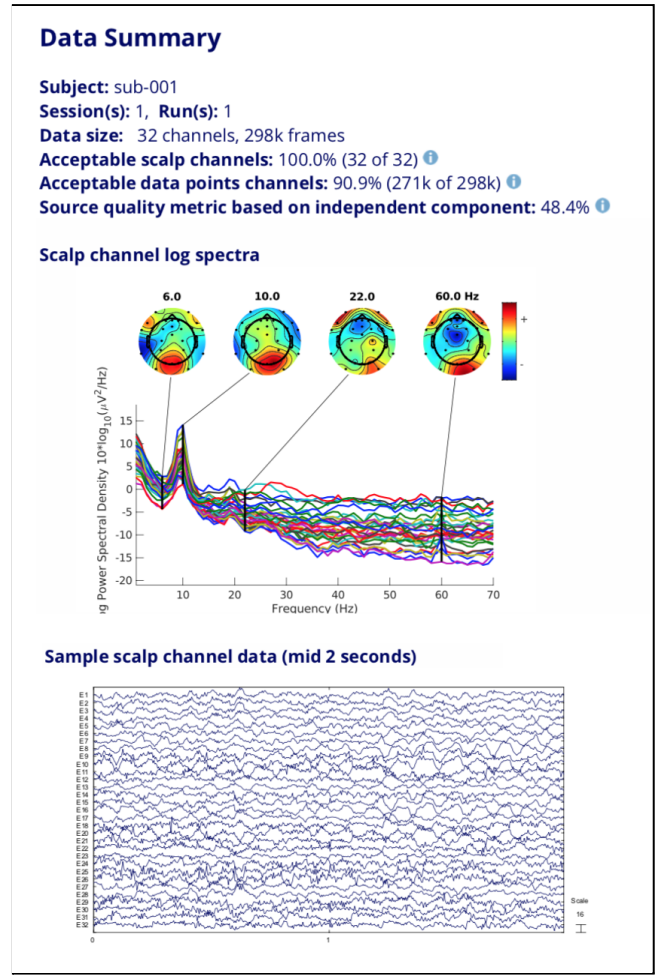

*Figure 4. Data quality information, channel log spectra, and raw data segment for a selected dataset.*

*E. NEMAR interface to the NeuroScience Gateway*

Since 2013, the Neuroscience Gateway (NSG) [9] has been serving the neuroscience community by providing easy access to many software and pipelines running primarily on high performance computing (HPC) resources provided by the Extreme Science and Engineering Discovery Environment (XSEDE) network that coordinates resources across the NSF-funded supercomputer centers.

NSG has made available high throughput computing (HTC) and academic cloud computing resources, and neuroscientists of all types have increasingly begun to take advantage of NSG and its easy-to-use interface for running tools and pipelines for processing their data. In addition to XSEDE HPC resources, implemented using Singularity, NSG uses backend XSEDE HPC, HTC and cloud resources and makes available a range of computational neuroscience tools including NEURON, PyNN, GENESIS, Brian2, NEST, MOOSE, NetPyNE, BMTK, HNN-Core etc. NSG has more recently made available environments widely used for human brain imaging research including Freesurfer, MATLAB, Python. In 2017, we ported to NSG our very widely used EEGLAB signal processing environment EEGLAB [11] and its many contributed tools and toolboxes, creating an Open EEGLAB Portal to high-performance data processing resources [10].

The NEMAR gateway project shares the same infrastructure as NSG, and NSG capabilities have been expanded to allow users to run data processing tools and pipelines on NEM data of their own or from the OpenNeuro archive. NEMAR already uses NSG for computing NEM data quality metrics (see III.D), and also allows users to run custom MATLAB and Python scripts on NSG (*nsgportal.org*) using NEMAR data (Fig. 5).

```
eeglab; % start EEGLAB and add paths
[~,EEG] = pop_importbids('/expanse/projects/nemar/openneuro/ds002691');

% duration and number of events for EEG dataset
duration = round([EEG.pnts])/EEG(1).srate);
nevents  = cellfun(@length, {EEG.event});

% display some statistics about the RAW EEG data
fid = fopen('results.txt', 'w'); % open result file
fprintf(fid,'BIDS dataset %s\n', bidsName);
fprintf(fid,'%d datasets\n', length(EEG));
fprintf(fid,'%d to %d channels\n', min([EEG.nbchan]), max([EEG.nbchan]));
fprintf(fid,'%d to %d seconds\n', min(duration), max(duration));
fprintf(fid,'%d to %d events\n', min(nevents), max(nevents));
fclose(fid); % close result file
```

*Figure 5. An example MATLAB script running on NSG using OpenNeuro / NEMAR dataset ds002691. The script, submitted by a user through the NSG interface, is executed on the Expanse supercomputer on which the NEMAR datasets are immediately accessible.*

## IV. Conclusion

The NEMAR portal (*nemar.org*) allows users to search and explore BIDS-formatted neuroelectromagnetic data submitted to OpenNeuro (*openneuro.org*) by viewing data quality metrics and visualized dataset information, and to process the selected data using the XSEDE high-performance resources in conjunction with The Neuroscience Gateway (*nsgportal.org*).

To support innovative NEMAR gateway data visualization and analysis tools, in future NEM data will be stored in both raw and source-resolved formats that will allow, for the first time, source-level data and data measure visualization for user inspection, search, analysis, and discovery within and also across studies using a common template brain model – obtaining results largely independent of scalp sensor geometry and modality that can be directly compared to results of other structural and functional neuroimaging research. This feature will also make possible data meta-analyses and large-scale data mining, as tools for this become available.

## Acknowledgment

The NEMAR project was supported by NIH grant R24MH120037. The OpenNeuro project is supported by NIH grant R24MH117179. We thank Erich Huebner and Claire Stirm, for their support on the NEMAR web interface.